\documentclass[11pt,twocolumn,a4paper]{article}
\usepackage[utf8]{inputenc}
\usepackage{xcolor}
\usepackage[colorinlistoftodos]{todonotes}
\usepackage{import}
\usepackage{listings}
\usepackage{paralist}
\usepackage{subcaption}

\usepackage[affil-it]{authblk}

\lstdefinelanguage{SPARQL}{%
        morestring=[b]",
        morestring=[s]{<}{>},
        morekeywords={SELECT,FROM,WHERE,DESCRIBE,ASK,CONSTRUCT,FILTER,GROUP,BY,prefix,.,?,;,\{,\},[,]},%
        otherkeywords={.,?,;,\{,\},[,]},%
        sensitive=false
}
\lstdefinelanguage{N3}{%
        morestring=[b]",
        morestring=[s]{<}{>},
        morecomment=[l]{\#},
        morekeywords={.,?,PREFIX,;,\{,\},@,\,[,]},
        otherkeywords={.,?,;,\,,\{,\},@,[,]},
        sensitive=false
}
\definecolor{mygray}{gray}{0.4}
\definecolor{spitfireorange}{HTML}{d45024}
\def \ACRONYMLong {SPARQL for Networks of Embedded Systems}
\def \ACRONYM {SNES}
\def \SNES {SNES}

\def \LINQLong {Linked Data In-Network Query Processor}
\def \LINQ {LINQ}

\def \SELDALong {SPARQL/Embedded Linked Data Adaptor }
\def \SELDA {SELDA}

\title{SPARQL for Networks of Embedded Systems}

\author{Dennis Boldt}
\affil{University of Lübeck\\Institute of Telematics\\Lübeck, Germany\\Email: boldt@itm.uni-luebeck.de}

\author{Henning Hasemann\\ Alexander Kröller}
\affil{
Technische Universität Braunschweig\\
Algorithms Group\\
Braunschweig, Germany\\
Email: \{h.hasemann,a.kroeller\}@tu-bs.de
}

\author{
Marcel Karnstedt\\Christian von der Weth
}
\affil{
Insight@NUI Galway\\
National University of Ireland, Galway\\
Email: marcel.karnstedt@deri.org\\
Email: christian.vonderweth@deri.org
}

\begin{document}
\maketitle

\begin{abstract} 
The Semantic Web (or \textit{Web of Data}) represents the successful efforts towards linking and sharing data over the Web. The cornerstones of the Web of Data are RDF as data format and SPARQL as de-facto standard query language. Recent trends show the evolution of the Web of Data towards the \textit{Web of Things}, integrating embedded devices and smart objects. Data stemming from such devices do not share a common format, making the integration and querying impossible. To overcome this problem, we present our approach to make embedded systems first-class citizens of the Web of Things. Our framework abstracts from individual deployments to represent them as common data sources in line with the ideas behind the Semantic Web. This includes the execution of arbitrary SPARQL queries over the data from a pool of embedded devices and/or external data sources. Handling verbose RDF data and executing SPARQL queries in an embedded network poses major challenges to minimize the involved processing and communication cost. We therefore present an in-network query processor aiming to push processing steps onto devices. We demonstrate the practical application and the potential benefits of our framework in a comprehensive evaluation using a real-world deployment and a range of SPARQL queries stemming from a common use case of the Web of Things.
\end{abstract}

\section{Introduction}
\label{sec:introduction}
Originally, the Web was designed for publishing documents, i.e., web pages, for human consumption, with no or only limited machine-readable access to the content. While platforms provide standardized APIs to expose their data, each platform uses its own data format to make the data available. As such, there is no unified way to access data across multiple platforms, thus forming a landscape of disconnect data silos. To remove the boundaries between these disparate data sources, the Linked Data concept~\cite{Bizer08LinkedData} provides guidelines on how to use Web technologies to publish and link data from different sources, thus forming the \textit{Semantic Web} or \textit{Web of Data}. Due to the application of standard technologies for the identification, representation and retrieval of data, the Web of Data can be accessed and explored in a unified way. Linked Data uses HTTP URIs for the identification of resources, where a resource refers to any real-world entity (e.g., person, object). The 
representation of information about a resource is done in RDF (Resource Description Framework). RDF is a data format using a simple subject-predicate-object triple pattern to make assertions (e.g., ``[Sensor A] [measures] [21.8C]''). Having the same URIs for the subject or object implicitly links triples, forming the so-called global RDF graph. The de-facto standard to retrieve of RDF data is the SPARQL query language, an SQL-like language optimized to access the RDF graph.

From the start, the data sources of the Web of Data were traditional online platforms. With the increasing omnipresence of embedded devices and networks, recent trends show the evolution towards the \textit{Web of Things}. The data stemming from these devices typically do not share a common format, making the integration and instant querying impossible. The goal is to apply the same Linked Data principles to the data to connect them into the global RDF graph, thus making them explorable and queryable as any other data source. In this paper, we focus on real-world embedded networks. We assume the common network architecture of a large set of devices accessed by a base station. Due to the resource-constrained nature of such networks (limited bandwidth and storage/processing capabilities) the application of the Linked Data principles poses two main challenges. Firstly, data stored in RDF tends to be rather verbose, potentially resulting in impractical large storage requirements. And secondly, the evaluation of 
SPARQL queries requires sophisticated optimization techniques to minimize the involved processing and communication costs.

To address these challenges, we present \textit{\ACRONYMLong} (\ACRONYM), a framework to make embedded networks first-class citizens of the Web of Things. It abstracts from individual deployments to represent them as common data sources in line with the Linked Data principles. This includes that our framework supports the execution of arbitrary SPARQL queries over the data from a pool of embedded devices and/or external data sources in the Web of Things. To the outside, \ACRONYM{} poses as a traditional SPARQL endpoint, accepting user queries and returning the results. In a nutshell, \ACRONYM{} analyses each query, identifying the subqueries that refer to data stored on the sensors and to data stored on external data sources. The in-network subqueries are constructed in such a way to push query processing steps onto the devices as much as possible to minimize the transferred data volume (between devices and the base station). Finally, \ACRONYM{} collects the partial results of all subqueries, integrating them to 
the final result, and returning it to the user. The main component of \ACRONYM{} is \textit{\LINQLong}\ (\LINQ) for the distribution and execution of SPARQL (sub)queries within an embedded network. We demonstrate the practical application our framework including LINQ in an evaluation using a real-world sensor deployment. We conduct our evaluation in the context of a typical Web of Things use cases from which we derive the set of SPARQL to be executed. Our results show that we can seamlessly integrate embedded networks into the Web of Things, while taking into account the resource-constrained nature of embedded environments.

Paper outline: 
Section~\ref{sec:relatedwork} reviews related work.
Section~\ref{sec:preliminaries} gives a brief introduction into RDF and SPARQL, and highlights the challenges of handling Linked Data on embedded systems.
Section~\ref{sec:usecase} outlines the main assumptions for the design of \ACRONYM{}.
Section~\ref{sec:architecture} presents the our framework in detail, describing \LINQ
and the accompanying SPARQL endpoint \SELDA.
Section~\ref{sec:evaluation} present the results of our evaluation and
Section~\ref{sec:conclusions} concludes the paper.

\section{Related work}
\label{sec:relatedwork}

Adopting common data management and data retrieval techniques for the Web of Things involves multiple challenges~\cite{James09ChallengesIoT}. For embedded devices to be useful without a supporting infrastructure, enabling a unified access and communication between devices, it is necessary to manage a certain amount of data on the device itself. Sadler and Martonosi have developed a database for embedded devices in Delay Tolerant Networks (DTN)~\cite{Sadler2007}. Distributed tuple spaces have been presented in the TeenyLIME\cite{Ceriotti2011,Costa2007} and Agilla~\cite{Fok2009} systems, which are available for TinyOS~\cite{springerlink:tinyos}. Tsiftes et al.~\cite{Tsiftes} have introduced a relational database for devices running the Contiki OS. All of these allow managing general-purpose data, but do not feature compact representation of RDF data. As RDF in its common 
serializations (such as RDF/XML\footnote{http://www.w3.org/TR/REC-rdf-syntax/} and N3\footnote{http://www.w3.org/TR/turtle/}) tends to be verbose, efficient compressing storage schemes are needed.

A lot of research has been done to support a more standardized access to sensor networks. \textsc{Cougar}~\cite{Yao:2002:CAI:601858.601861}, \textsc{TinyDB}~\cite{Madden:2005:TAQ:1061318.1061322}, \textsc{SNEE}~\cite{Galpin:2011:SQP:1921816.1921819} and related approaches (e.g.,~\cite{Rohm07OnTheIntegration,Lee07EfficientTime}) consider a sensor network like a distributed, relational database. With sensor data being viewed as a virtual database table, these systems provide SQL-like query languages that are able to collect, filter, and display data from sensor networks. The goal is to abstract from tasks such as sensing, data transmission and data merging/aggregation in resource-constraint environments. Compared to processing SQL-like queries over relational n-tuples, the processing of SPARQL queries over RDF triples results in much more join-heavy query processing.

In-network solutions to efficiently process joins in resource-constraint networks has been studied extensively~\cite{Kang13InNetwork}. The goal is to minimize the data volume transferred between devices to calculate join results.
\textsc{REED}~\cite{Abadi:2005:RRE:1083592.1083681} addresses joins between an external relation and the sensor relation, pushing the relevant tuples of the external relation into the network. Approaches such as ~\cite{Coman07OnJoinLocation,Yang:2007:IEM:1247480.1247538,Zhou06InNetworkJoin} implement a \textit{semijoin}. Here, in a first step, only the minimal sets of attributes of tuples to calculate a join are communicated between devices. In the second step, only relevant source tuples need to be transferred to complete the join result. \textsc{INJECT}~\cite{Min20113443} uses Bloom filters to identify the potentially relevant data before transferring the actual data. \cite{Lai:2008:PPE:1594696.1594703,Pandit06CommEfficient} implement a \textit{hash join}, partitioning the virtual sensor data relations into buckets which are replaced inside the network to perform the join operation. All these works assume relational n-tuples. Thus, the number of join operations is rather low. Furthermore, the proposed solution have been evaluated using theoretic analyses or by means of simulations, not using real-world deployments.
\\
\\
In contrast to existing efforts, this paper addresses the evaluation of SPARQL queries over RDF data in networks of embedded systems to make them first-class citizens of the Web of Things. Similar to existing embedded query engines, we try to push processing steps into the network down onto devices. The goal is to minimize the volume of data transferred between devices. To deal with typically high number of join operations when processing SPARQL queries, we assume a graph-based description of the data stored on the devices to be available. This data description allows us to decide where to execute a specific join operation.

\section{Linked Data Management}
\label{sec:preliminaries}
In the following, we first give a brief overview to RDF as well as SPARQL, and then outline the major challenges towards integrating embedded networks into the Web of Things.

\subsection{RDF -- Resource Description Framework}
\begin{figure}[htp]
 \centering
 \includegraphics[width=0.40\textwidth]{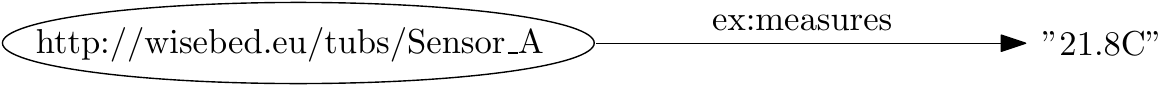} 
 \caption{Graph representation of RDF triple}
 \label{fig:exp-rdf-triple}
\end{figure}

RDF is a simple data format to express assertions about resource. A resource can refer to any real-world object or person. Consider the following example: “Sensor A measures 21.8C.” Each statement is represented in the form of a triple that links a subject (“Sensor A”), a predicate (“measures”), and an object (“21.8C”). The subject is the resource that is described. The predicate is a term used to describe or modify some aspect of the subject. The object is the ``target'' or ``value'' of the triple. It can be another resource or just a literal value such as a number or word. To publish data on the Web, resources need to be uniquely identified. In RDF, resources are represented by Uniform Resource Identifiers (URIs). The typical way to represent a RDF triple is a graph, with the subject and object being nodes and the predicate a directed edge from subject to object; see Figure~\ref{fig:exp-rdf-triple} for an example. 

\begin{figure}[htp]
 \centering
 \includegraphics[width=0.40\textwidth]{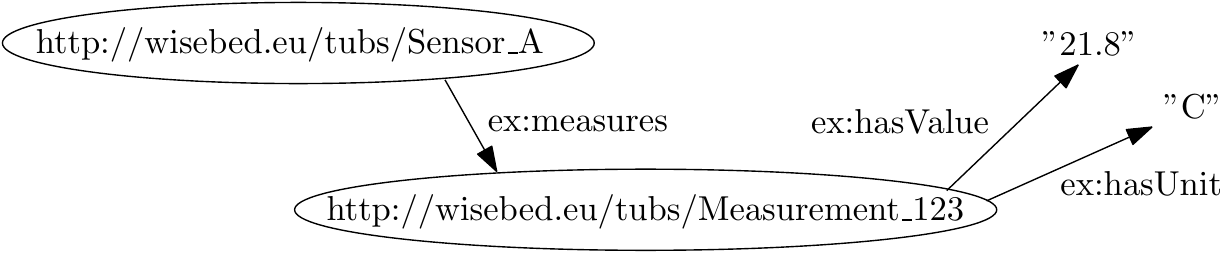} 
 \caption{A simple RDF graph}
 \label{fig:exp-rdf-graph}
\end{figure}

Since objects can also be resources with predicates and objects on their own, single triples are connected to a so-called RDF graph. An RDF graph is a labelled and directed graph. As illustration, we extend the previous example, replacing the literal ``21.8C'' by a resource ``Measurement''. This resource has two predicates assigning a unit and the actual value to the measurement; see Figure~\ref{fig:exp-rdf-graph}.

Using unique URIs and the simple data format of RDF makes it easy to publish data on the Web in a unified way, and allows for a straightforward integration of data from different sources. For example, a RDF triple stored in one data source can have an object with a URI referring to resource for which more information is stored on another data source.

\subsection{SPARQL: Querying RDF Data}
SPARQL (SPARQL Protocol and RDF Query Language) is the de-facto standard query language to retrieve RDF data. The structure of SPARQL queries is very similar to SQL. The most basic construct of a SPARQL query is a \textit{basic graph pattern} (BGP). Such a pattern  is very similar to an RDF triple with the exception that the subject, predicate or object may be variables. A BGP represents a basic triple filter, matching all RDF triples whose terms (e.g., URI or literal) may be substituted for the variables. For example, the BGP \textit{?sensor ex:attached\_to ex:System\_S1} matches all triples with predicate \textit{ex:attached\_to} and object \textit{ex:System\_S1}. Using same variable names also allows combining multiple graph patterns. The following SPARQL query returns all sensors attached to systems located in the Building A:

\begin{lstlisting}[language=SPARQL]
PREFIX ex: <http://example.org/>
SELECT ?sensor 
FROM <http://example.org/>
WHERE {
  ?sensor ex:attached_to ?system .
  ?sensor ex:located_in ex:Building_A .
}
\end{lstlisting}

Besides BGPs, the SPARQL also supports further concepts known from SQL, such as the sorting (ORDER BY), the limitation of result sets (LIMIT, OFFSET), the elimination of duplicates (DISTINCT), the formulation of conditions over the value of variables (FILTER), grouping (GROUP BY) including the formulation of conditions on groups (HAVING), the aggregation of results (COUNT, SUM, MIN, MAX, AVG, etc.), and others. As an illustration, we extend previous query to retrieve the number of sensors in Building A that measure more than 20 degrees Celsius for each system in the building:

\begin{lstlisting}[language=SPARQL]
PREFIX ex: <http://example.org/>
SELECT ?system COUNT(*) AS ?count 
FROM <http://example.org/>
WHERE {
  ?sensor ex:attached_to ?system .
  ?sensor ex:located_in ex:Building_A .
  ?sensor ex:measures ?temp .
  FILTER (?temp > '20') .
}
GROUP BY ?system
\end{lstlisting}

\subsection{Challenges}
Integrating embedded devices and networks into the Web of Things faces many practical challenges. These challenges derive from the fact that RDF and SPARQL are not well-suited for an direct application in resource-constraint environments.
\\
\\
\textit{Verbosity of RDF.}
Particularly because of the use of typically long URIs, RDF is a very verbose data format. While this is almost a non-issues for traditional data sources on the Web (e.g., Web servers), a direct application of RDF to embedded environments is not practical given the typically very limited available resources in terms of local storage, bandwidth, and battery lifetime of individual devices and networks of such devices. Thus, instead of handling URIs as plain string, we apply hashing techniques in combination with dictionaries to store, process and transfer RDF data.
\\
\\
\textit{Join-heavy query processing.}
Most SPARQL queries include multiple basic graph pattern linked via shared variables. Each shared variable results in a join operation, with matching triples potentially residing on different data sources. Thus, in principle, for each join operation, relevant triples may reside on different devices. For any real-world deployment, the distribution of required triples among all devices or to the base station is not practical. We therefore propose a simple data model describing the structure of the RDF data locally stored on embedded devices. With that description we can identify join operations derived from SPARQL where all potentially relevant triples reside on the same sensor node. Join operations that require data coming from different devices are processed on the base station.
\\
\\
\textit{SPARQL query processing.}
The efficient evaluation of SPARQL queries is a non-trivial task in resource-constraint environment. Naive approaches such as forwarding all to the base station typically involve unnecessarily high communication costs since most triples do not contribute to a query result. We therefore aim to push query operations into the network. For this, our framework analyzes queries to distinguish between three different categories of operations:
(a) operations that can directly be executed on the RDF data locally stored on devices, such as the evaluation of BGPs, simple filter operations and local joins, 
(b) operations that can be executed within the network, i.e., over intermediate results that are forwarded towards the base station (e.g., aggregations), and
(c) operations that need to be executed on the base station, such as joins over non-local data or sorting.

\section{Use case scenario}
\label{sec:usecase}

We have to face the fundamental challenges when implementing data management and processing techniques running over devices with limited capabilities. Firstly, the size of the data handled needs to be minimized. And secondly, the low-level programming of resource-constraint devices benefits from handling data structures of fixed sizes to avoid dynamic memory allocation and management (negatively affecting the overall code size, the efforts for the programmer and the probability of bugs). Based on these basic goals, we make the following assumptions affecting the design of \ACRONYM{}:
\\
\indent \textit{(1) Minimal device capabilities.}
We assume deployments comprising severely resource-constrained embedded devices with regard to computing power, local memory and bandwidth. 
\\
\indent \textit{(2) Powerful base station.} We assume the common network architecture with a base station that acts as endpoint for communicating with devices. Compared to individual devices, the base station has virtually unlimited resources.
\\
\indent \textit{(3) Local-scale sensor deployments.} We assume deployments of a local scale with a maximum of only a few hundred devices, confined to a well-specified area (e.g., a building). The characteristics (e.g., routing paths) of the deployment do not frequently change over time. All devices devices are of the same or similar type with the same or similar capabilities.
\\
\indent \textit{(4) Maximum query complexity.} We assume that user queries cannot be ``too'' complex, with no query having of more than 256 operators, and no (intermediate) result row being longer than 16 columns.

\section{Architecture}
\label{sec:architecture}

Our system is split into two physical components: The \SELDALong{} (\SELDA{}) and \LINQLong{} (\LINQ{}). The former represents the interface of the System towards the Web by providing an HTTP SPARQL endpoint. Furthermore it fulfills the task of parsing, splitting and optimizing SPARQL queries into several subqueries. These subqueries are either answerable completely in the Web, or completely in the embedded network and are sent by \SELDA{} to their according destination. While external subqueries are sent to another HTTP SPARQL endpoint or answered using \SELDA{}s local tuple store, subqueries are sent to the embedded network to be processed by \LINQ{}.

The following section will give an overview over the workings of \LINQ{} and how data is processed within the embedded network. Then we will give an overview on some design decisions in terms of query processing in \SELDA{}.

\begin{figure}[h!]
	\centering
	\includegraphics[width=0.45\textwidth]{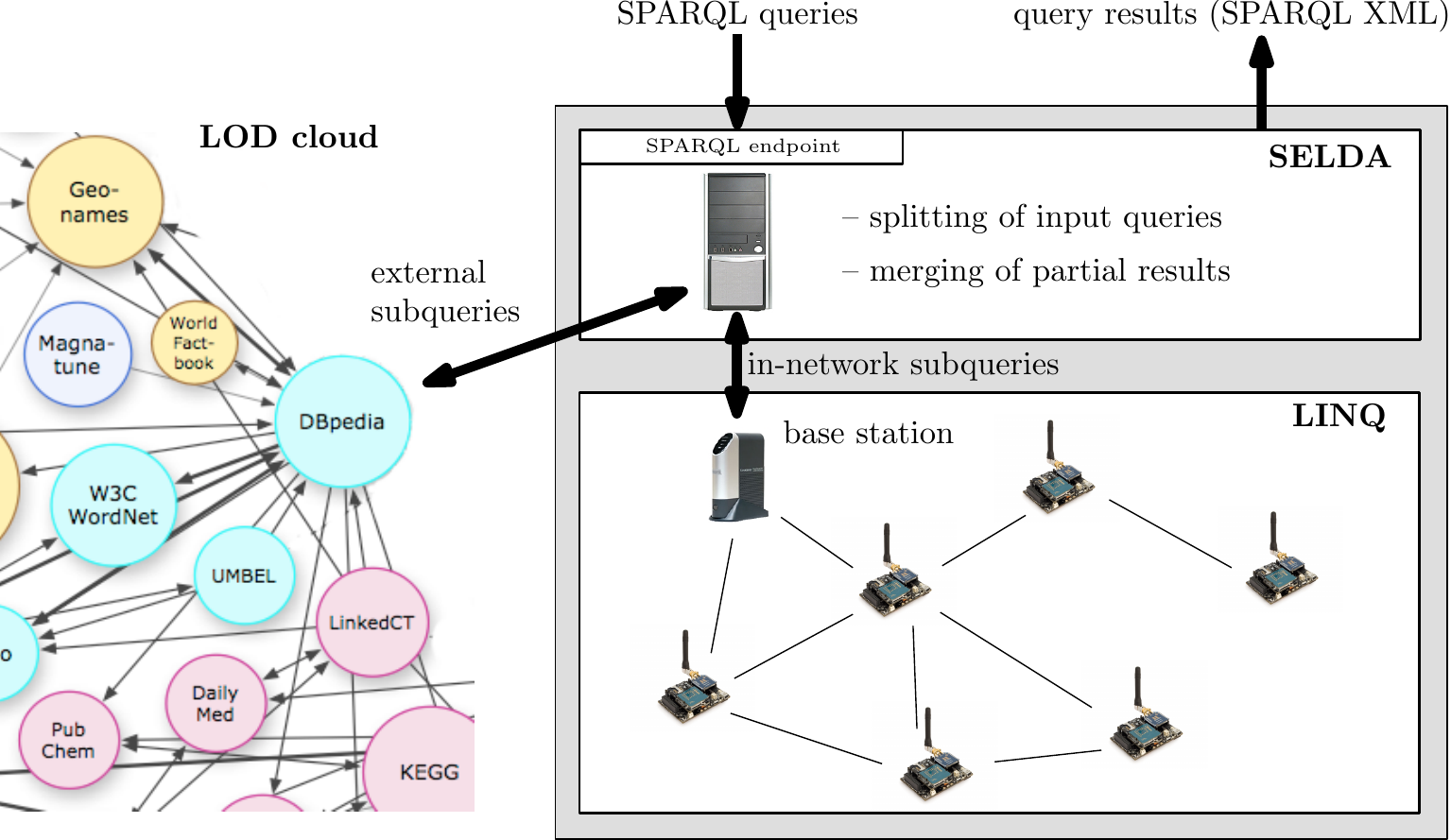}
	\caption{The architecture of \ACRONYM{}.}
\end{figure}

\subsection{Data Representation and Storage}

\newcommand{\xxx}{\,(\textasteriskcentered)}
\begin{figure}
	\centering
	\footnotesize
	\begin{tabular}{lclccl}
\hline \hline
Hash Function &  Collisions & Elems/hash \\
 & & \multicolumn{1}{c}{$\sigma^2$} \\
\hline
Additive (K\& R)      & 3267131 & 7.99e-01 \\
Bernstein (sum)       & 1959    & 7.73e-04 \\
Bernstein (xor)       & 1219    & 7.73e-04 \\
ELF                   & 170101  & 9.95e-04 \\
FNV1                  & 1303    & 7.73e-04 \\
FNV1a                 & 1569    & 7.73e-04 \\
Jenkins Lookup 2      & 1219    & 7.73e-04 \\
Jenkins Lookup 3      & 1269    & 7.73e-04 \\
Jenkins One-At-A-Time & 1392    & 7.73e-04 \\
Larson                & 1088    & 7.72e-04 \\
Novak                 & 1060380 & 3.01e-03 \\
SDBM                  & 1070    & 7.72e-04\xxx \\
\hline \hline
	\end{tabular}
	
	\caption{%
		Hash values for different 32-bit hash functions for each of the 3,320,193 different tuple
		elements found in the ``datahub'' data set of the Billion Triple
		Challenge 2012.
		In addition to the number of collisions, the variance of the number of elements per hash value gives an indication of the quality of hash-value distribution (low values are better).
		\xxx \, marks the best results wrt.\ runtime and variance in each group.
	}
	\label{fig:hashes}	
\end{figure}

As RDF data tends to be very verbose, an efficient representation and storage of data elements on resource-constrained devices is crucial. We distinguish between data representation for storage on the device and query processing / communication. For storage on devices, we use the Wiselib TupleStore~\cite{tuplestore}, a database for managing RDF on resource-constrained embedded devices. The TupleStore stores the tuple elements encoded as strings in a special compressing \textit{Dictionary} which allows them to be identified by locally generated keys. Tuples are then represented compactly
with sets of fixed-size dictionary keys. A user of the TupleStore can choose whether to iterate over the tuples as tuples of strings or tuples of dictionary keys.

While in the common encodings all elements of an RDF document are strings, other describe numeric values (such as sensor measurements), usually encoded decimal. It seems, however, that there is a natural demand to an embedded network query engine to be able to process numeric data for aggregation and other operations. Also, for in-network query processing, communication of tuple elements have to be exchangeable and comparable across devices. Thus, during query processing, we consider three different data types:

\textit{INTEGER.}
We represent integer values in the usual 32 bit integers on all platforms, as this allows for straightforward and well-established implementations of the usual operations and conversion from and to strings.

\textit{FLOAT.}
Similarly, fractional values are represented as 32 bit IEEE floating point values for straightforward conversion and utilization of hardware floating point units which are available even on many constrained devices.

\textit{STRING.}
Strings represent the most common data type in RDF data and are variable-sized and in RDF not uncommonly considerably long. Especially in distributed query processing of RDF data, likely many devices will output the same string elements as part of intermediate results even although it may be discarded (due to a failed match) at a later point. In order to avoid unnecessary sending of strings, for query processing we substitute them with 32 bit hash values. This allows sufficiently reliable equality comparisons for query operators while keeping the elements at a constant and small size. Figure~\ref{fig:hashes} shows an investigation of the collision properties of some hash functions on RDF data.

\subsection{Communication Infrastructure}
The communication infrastructure for in-network query processing is split into several layers: On the lowest layer, a routing tree, rooted at the device running \SELDA{}, cares for message delivery. Using the abstraction mechanisms of the Wiselib, any routing tree can be used from a tree defined by a flooding of queries to complex algorithms that maintain, e.g., a self-stabilizing tree in conjunction with neighborhood discovery systems. Additional properties like routing overlays or reliable transport mechanisms can be installed either as an adaptor to the operating system radio interface or as part of the tree implementation using Wiselibs powerful component stacking mechanisms. Thus, our communication implementation can focus on the essential message exchange and does not need to be concerned with details of message delivery.

Basing on the chosen tree, the following communication components exist:

\textit{String Requests.} The String Request Processor is responsible for resolving requests from \SELDA{} for strings belonging to hash values. In order to be able to answer these string requests, the string request processor maintains a cache of assignments of hash values to dictionary keys. A device receiving a string request searches its cache for the according hash value / string pair,
extending the search to an iteration over the dictionary if necessary. If successful, the according string is send upwards along the tree to \SELDA{} for processing. Otherwise, the request is forwarded to the devices' children.

\textit{Query Distribution.} When issuing a query, \SELDA{} will send it to all its children, splitting it into multiple messages when necessary, such
that each message can be interpreted on its own (i.e., it only contains complete components such as query operators). The Query Distribution
component on each receiving device will process the (partial) message and at the same time ensure it is forwarded to all children. Queries contain a query id, a set of operators (each with an id and several parameters) and a lifetime that determines how long devices should keep processing the query before discarding it.

\textit{Intermediate Results.} As we will discuss in detail below, two operators can communicate intermediate results through the network: \textit{Collect} and \textit{Aggregate}. When the Collect operator running on a device sends out a tuple of intermediate results, it will be forwarded upwards along the tree, such that it finally arrives at \SELDA{}.  Tuples produced by the \textit{Aggregate} operator are sent upwards along the tree as well. However, devices receiving those tuples will not simply forward it, but rather pass it on as input to the local Aggregate instance for aggregation. All intermediate results are tagged with the id of the generating query and operator such that they can always be delegated to the correct operator instantiation.

\subsection{In-Network Query Processing}

\subsubsection{Basic Design Decisions}

\LINQ{} processes queries formulated as trees of operators in a distributed fashion within the embedded network. As these embedded devices are commonly battery powered and communication is typically expensive, special care must be taken to choose an efficient set of operators and to encode them compactly.

\textit{Stateless vs. stateful operators.} Stateful operators require additional information, i.e., the operator state, to process an individual row. E.g., the state of a join comprises all input rows that have been processed so far by the operator. For aggregation operators (SUM, COUNT, AVG, MIN, MAX) the operator state basically includes the current intermediate processing result. Stateless operators only
require constant data to process a row. For example, the filter condition of a selection operator is confined to an attribute value of a row. 

\textit{Blocking vs. non-blocking operators.} Blocking operators require to process all input rows before returning the final results. For example, aggregation is blocking since all rows need to be examined before the specified maximum, minimum, average, etc. is available. Operators like selection or projection are non-blocking operators, allowing the return of individual rows once they have been processed. 

For embedded networks, given their distributed nature and the limited capabilities of devices, stateless and non-blocking query operators are most suitable. Firstly, relying on stateful operators potentially affects result of queries in case of device failures. Also, to keep the operator state on devices requires additional storage, which can be significant in case of joins. Secondly, given the latency due to transferring data, non-blocking result rows can immediately be processed by the next operator according to the query plan.
Blocking operators need to know when the last input row has arrived. In distributed settings, this information is typically not available. 


\subsubsection{Operators}
\label{sec:operators}

\paragraph{Graph Pattern Selection} The Graph Pattern Selection (GPS) allows simple selection of tuples based on equality comparison to constants. This reflects the very common operation that a basic graph pattern (BGP) in a SPARQL query expresses, namely to select tuples with certain known values for subject, predicate or object, while leaving some positions variable. The Graph Pattern Selection operator can only be used as a leaf in the local operator tree. Instead of other operators, the GPS receives its data from the TupleStore.

\paragraph{Complex Selection} The complex selection -- in contrast to GPS -- can be applied to the results of other operators and provides selecting tuples based on the usual comparison operators ($<, \le, =, \neq, \ge, >$). Comparisons can be executed on any combination of attributes and constants which are provided as additional parameters to the operator.

\paragraph{Simple Local Join} Commonly, SPARQL queries include several joins that connect the individual graph patters by variables. Very often, these joins can be evaluated locally, as all resources are fully described on a single device. In order to cover this common case, we added the Simple Local Join operator (SLJ).  SLJ has 2 attribute
indices as parameters. SLJ receives tuples from 2 children in the operator tree and outputs a tuple for each combination of input tuples that compares equal for the given attributes.

\paragraph{Collect}
The Collect operator sends out all tuples it receives from child operators along the established routing tree towards \SELDA{} for further processing. As it does not produce any local output, it can only be used as the root of the operator tree of a device-local query part.

\paragraph{Aggregation} In addition to local child operators, aggregation also
receives tuples from its routing tree children. The operator internally keeps
track of the aggregation state and updates it upon reception of tuples either
from child operators or child devices. Within fixed time intervals, if the
aggregation state has changed within the interval, the current state is sent
to the parent device in the tree. This way, each device maintains a continuously
updated aggregation of its subtree. Our implementation supports the aggregation 
functions SUM, COUNT, AVG, MIN and MAX.

\paragraph{Projection Bitmask} \label{sec:projection} In order to be able to
execute queries on embedded devices with potentially very limited RAM, it is
not only essential to encode queries and data efficiently, but also to avoid
processing of intermediate data that is known to be discarded at a later
point. We observe, that most relational operations in their usual
form output some attributes that are not relevant for further processing of
the query. For example, after selection on attribute equality, at least one
attribute contains redundant information. Thus we
designed the operators in \LINQ{} to always include a projection, defined
by a compact bit mask of 4 bytes. This has the following advantages over a
dedicated projection operator: 

\begin{compactenum}[1.] \item It avoids 7 bytes
of space usage for each potential dedicated projection operator instance at
the expense of adding its 4 byte bit mask to every other operator. As we
believe the projection operator to be very common, this representation is very
likely the more compact one.

	\item The projection bit mask also carries type information for all
attributes processed by the operator and can thus ensure correct handling of
the data. As with this method the relevant output attributes are known to the
operator, it can apply further optimizations. In the example of a join
operation, the temporary table being allocated thus does not need to store
attributes that will be discarded by projection.
\end{compactenum}

\subsubsection{Query Representation} \LINQ{} queries are generated by \SELDA{}
distributed to all devices in the embedded network to be executed on all devices
in parallel. These queries are composed of the following components: 

\begin{compactenum}[1.] \item A unique query ID.  \item The number
of operators this query contains.  \item The query lifetime.  \item A set of
operator definitions.
\end{compactenum} 

Operators, the main element of our in-network queries, each
consist of 

\begin{compactenum}[1.] \item A query-local operator id.  \item The
type of the operator as described in Section~\ref{sec:operators}.  \item The
id of the parent operator or \texttt{NULL}.  \item The projection bitmask as
described in Section~\ref{sec:projection}.  \item Operator-type specific
parameters of variable size. \end{compactenum}

Special care has been taken to implement a compact encoding of operator
descriptions (typically encoded in 20 bytes), such that operators do not need to be split into
multiple messages. If a query should exceed the maximum transfer unit (MTU)
and cannot fit in a single message, \SELDA{} will split it at operator
boundaries such that each message can be directly interpreted by the receiver
and thus a memory intensive assembly messages is not necessary.

\begin{figure}
	\centering
	\def\svgwidth{.5\columnwidth}
	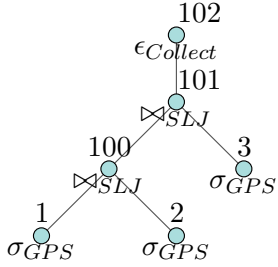
	\caption{Sample operator tree.}
	\label{fig:optree}
	\vskip -2em
\end{figure}

For processing, operators are instantiated, such they can allocate memory for
temporary storage and send tuples to their parent operator using a redirecting
\textit{push()} method. This way, operator instances form a tree defined by
their parent relationships, shown in Figure~\ref{fig:optree}. In
order for the query result to be sent back to \SELDA{}, the root of this tree
has to be either a Collect or an Aggregate operator.
Operators are executed in ascending order of their ID, such that in this
example, the GPS with the IDs 1 to 3 will be executed in
that order. Operators other than the GPS do not implement an execute method
and thus will not be executed. They are, implicitly
activated by child operators pushing results to them, such as a SLJ caching
tuples that arrive from the left child or possibly sending out join results
triggered by reception of tuples from the right child. In order to allow
possibly allocated memory to be freed as soon as possible, once an operator
has finished pushing out tuples, it will send a special \texttt{NULL} tuple to
inform its parent that no further tuples will be sent.

\subsection{Query Planning} Due to the universal applicability of RDF, there
is, no limitation on what sets of triples are stored on the embedded devices.
As consequence, triples that need to be joined according to graph patterns
sharing the same variable(s) may reside on different devices or even on
resources outside the sensor network. Assuming the worst case this would mean
that all pattern joins would need to be implemented as base station joins,
degenerating \LINQ{} to shipping all relevant RDF triples to the base station.
Often joins can be executed locally on the embedded device such that
it is not necessary to conduct such an exhaustive tuple collection.

For other query parts it might be clear that the devices cannot
old matching tuples as the according knowledge domain is not covered
by any of them. Accordingly, a SPARQL query may be logically
split into several parts:

	\textit{Device-local.} Parts of the query might be answerable locally on
each device before any communication of query results.
This trades (relatively inexpensive) local communication for costly radio
transmission of tuples.

	\textit{Web.} Some parts of the query refer only to external data sources
and are thus answerable by sending subqueries to other SPARQL endpoints. 
This part does not stress the embedded network at all but can be executed 
completely on high-performance machines in the World Wide Web.

	\textit{In-Network.} Query parts that demand for aggregation of values from
different devices, or joins that combine tuples from different devices require
communication across the network and can thus be neither executed completely
locally nor completely by \SELDA{}. In-network subqueries are represented by
the two in-network operators \textit{Collect} and \textit{Aggregate}.

	\textit{Base-Station-local.} The decomposition of incoming SPARQL queries
into the subqueries for the three query targets described above must naturally
happen and the base station. During this splitting process, \SELDA{} records
the points at which the initial operator graph has been partitioned.
Furthermore, finalizing operations such as sorting or aggregation are
recorded. After the results for the several partial queries have been
collected, \SELDA{} joins them according to the recorded split points using an
in-memory relational database and applies the final operations.

\subsection{Data Model}

\begin{figure}
	\centering
	\includegraphics[width=.8\columnwidth]{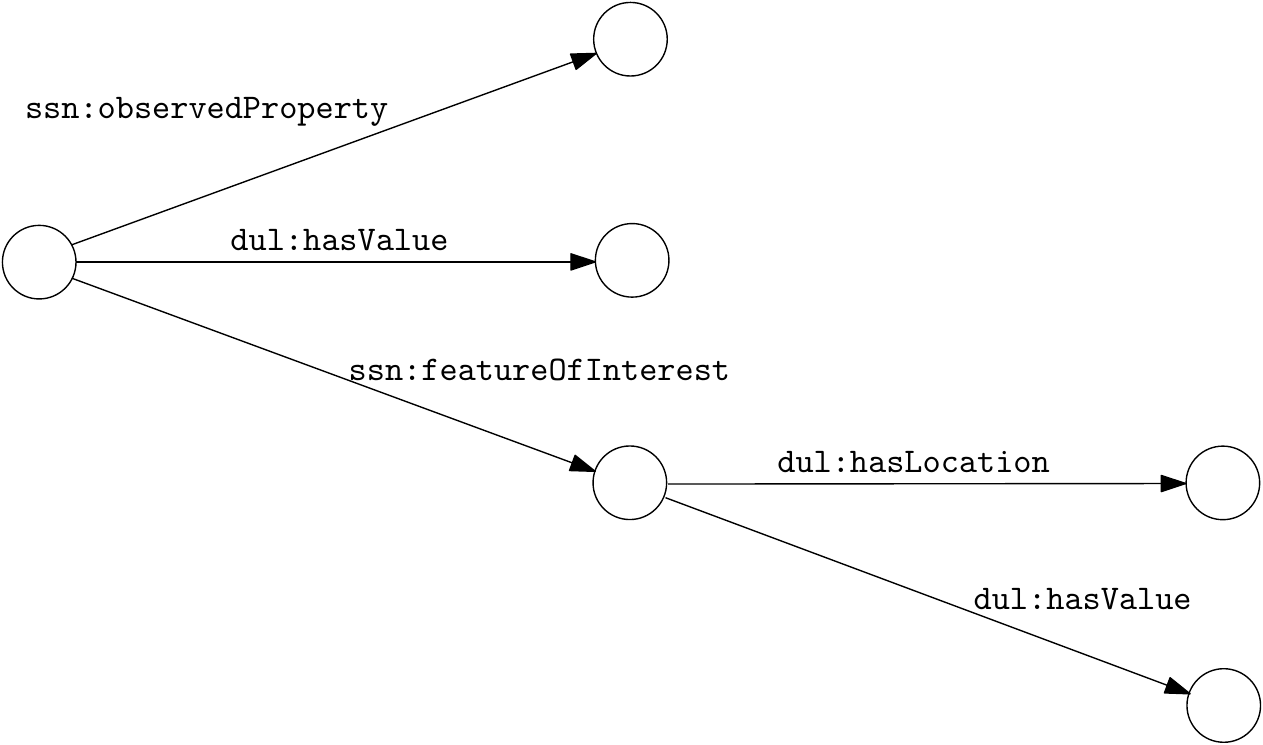}
	\caption{Example data model for locally answerable subqueries.}
	\label{fig:data_model}
	\vskip -2em
\end{figure}

In order to decide how to split an incoming query and where to execute which
part, a data model describes the knowledge of the embedded network. This
model is defined by a directed graph $G$ in which RDF subjects and objects are
represented by devices and predicates by labelled edges. Devices are unlabelled,
that is, they can match any resource, literal or variable.  Edges are labelled
such that they can only match a single predicate. A graph pattern $(S P O)$
with $S$ and/or $O$ being  potentially variables and the rest URIs or literals
is then considered answerable by an embedded device locally if an edge
matching $P$ can be found in the model. A join of two graph patterns, e.g.,
$(S P_1 X)$ and $(X P_2 O)$ is consequently considered answerable locally if a
directed path with edges $P_1, P_2$ exists in the model. This notion is
directly extended to general subgraphs: Upon receiving a SPARQL query with
pattern graph $Q$ for splitting, \SELDA{} computes the maximal common
subgraphs of $G$ and $Q$, these must be locally answerable by every device, thus
\SELDA{} models joins within these graphs as simple local joins.


\subsection{Discussion}



\textit{Heterogeneity.} The Wiselib provides lightweight abstractions on top of several operating
systems such as Contiki, TinyOS and iSense. This makes it easily possible to compile \LINQ{} for all those
platforms. Due to the large variety in terms of available resources among embedded
devices this alone would hardly be enough: While some devices feature large
block storage devices (e.g. SD cards), large RAM and powerful processors,
others may have limited RAM, be battery powered
and not even provide dynamic memory allocation. Depending on the deployment it
might also be necessary to implement additional transportation mechanisms such
as reliable or encrypted communication. Thanks to the
modular approach that is followed by both the Wiselib and the
Wiselib TupleStore, these components (Radio communication, storage devices and used data structures) can be exchanged, adapted and
extended with a few lines of C++ template instantiations.  This way, with
minor efforts, we can add an encryption layer or change the system to work on tuples stored on a SD 
card instead of in RAM.  These abstraction decisions happen -- in contrast to
mechanisms relying on virtual inheritance -- at compile time and thus, do not
have an impact on performance (such as code size, or indirect calls).

\textit{Generalizability.} \LINQ{} follows the same principles: By the use of template instantiations the
internal data structures (e.g., those used for storing the operator tree) can
be easily exchanged, so a user can decide whether to use statically allocated
structures or dynamic ones. Moreover, the components for query distribution,
resolving hashes to strings, processing queries and the individual operators
are exchangeable. Additionally, care has been taken to separate the
descriptions of operators (as exchanged via radio) from the operator
implementations. The current restrictions (max. 16 attributes per
intermediate result, max. 256 operators, etc.) can thus be raised just by
implementing new descriptions. While the Wiselib TupleStore is tailored
specifically for the storage of RDF it can also be used to store other types
of data such as numeric values which in turn could be processed by \LINQ{} by
adapting the GPS implementation.

A few core design decisions remain inherent to the system and cannot be raised
without affecting multiple components:

\textit{All operators include projection information.} This is crucial for
determining how to handle received attributes (i.e., as what type) and thus
also for string $\leftrightarrow$ hash translation.

\textit{The operator tree is executed using a 'push' approach.} While also a
\textit{pull} method might be thinkable -- that is, executing the tree root
operator which will request child operators for tuples -- the assumption of a
push-based mechanism is inherent to the system and can not easily be changed.

The current implementation of \SELDA{} meets the requirements of our use
case setting. \SELDA is responsible for splitting incoming user queries
into subqueries and forwarding them to \LINQ{} and external data sources. The
splitting of user queries as well as the assembly of subqueries requires the
parsing of user queries. As a declarative language, given the users multiple
ways to formulate a query, SPARQL features are rich syntax. Moreover, compared
to SQL, SPARQL is still a rather new standard. As such, the SPARQL syntax and
grammar is still evolving with new constructs continuously being proposed and
implemented. The focus of this paper is on \LINQ{} and the proof-of-concept
implementation of \SNES{} to showcase the direct integration of embedded
systems into the Web of Things. Thus, the SPARQL parser of \SELDA{} covers the
most common language constructs as needed for the queries for our use case.
Adopting our framework would most likely require to extend the parser in order
to support the queries required by the new uses case or application setting.

\section{Evaluation}
\label{sec:evaluation}


\subsection{Features and Query Semantics}



\begin{figure}
	\centering

		\begin{lstlisting}[language=SPARQL]
SELECT ?time ?description WHERE {
	?node :observed ?condition .
	?condition :hasSeverity :Critical .
	?condition :hasTimeStamp ?time .
	?condition :hasDescription ?description .
}
\end{lstlisting}

	\caption{SPARQL query requesting all conditions that classify as
``critical''.}
	\label{fig:graph-queries}
	\vskip -1em
\end{figure}

We consider a setup where embedded devices monitor several household devices by, e.g., observing their energy consumption, temperature or
physical integrity. An application running on the embedded device that -- thanks to information locally stored in RDF -- can turn these measurements into time-stamped \textit{conditions}, that
describe whether the observed device is working, switched on and so forth.
This allows to discard the individual, detailed measurements and only retain
the important, abstract device state.
When a user wants to find all critical situations that have been observed,
he/she might issue a SPARQL query as shown in Figure~\ref{fig:graph-queries}.
Critical conditions might have different meanings depending on the device and
its operations, ranging from inefficient operation to physical damage on the
device, thus we also query for a short, human readable description of the
condition. 
\SNES{} ensures that the critical conditions can be found locally on each embedded device, such that only information about the critical conditions ever has to be sent through the network.

\subsection{Experimental results}

\begin{figure}
	\centering
	\includegraphics[width=.8\columnwidth]{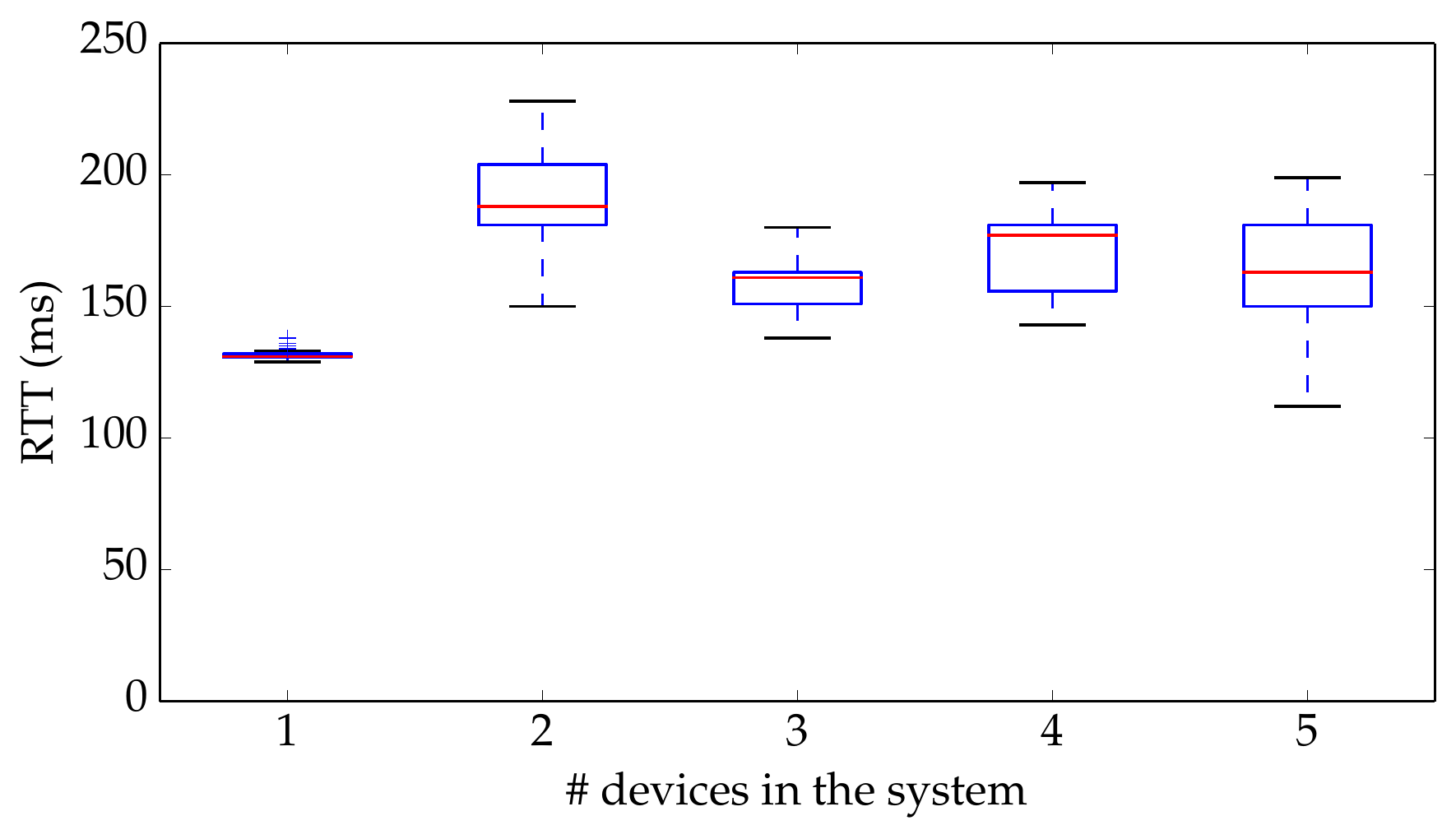}
	\caption{Distributions of response times (ms, y-axis), depending on the number of
sensors present in the system (x-axis).}
	\label{fig:timings}
	\vskip -1em
\end{figure}

For the experiments \SNES{} uses the public endpoint of
DBpedia\footnote{http://dbpedia.org/sparql} as an external data source and
an embedded network of Coalesenses\footnote{http://www.coalesenses.com} iSense sensors. We executed 100 queries against 
the SPARQL endpoint provided by \SNES{}, with one to five sensors present in
the network.
The subqueries were identified by our framework as described before.
Thus, one subquery was send to the external data sources and one subquery to \LINQ{}. 
The response time by the external endpoint was 190ms on average, 
Figure~\ref{fig:timings} shows the round trip times to the embedded devices, including the query processing
on each sensor.

\section{Conclusion} \label{sec:conclusions} With the Web of Things accessing publicly available data
from Web resources as well as embedded systems in a unified allow for novel types of services on the Web. The
established technologies of the Web of Data, however, were not designed to be
applied in resource-constraint embedded environments. Handling verbose RDF
data and executing SPARQL queries requires optimized data structures
and algorithms on both the storage and in-network data processing level.
\ACRONYM{} provides a novel solution, featuring a space-optimized data
structure for storing RDF on devices and in-network SPARQL query processor,
allowing us to directly plug into the Web of Things. We evaluated \ACRONYM{}
in a practical setting using a real-world deployment, confirming the
feasibility of our approach to make the Web of Things real.  In our on-going
work, we plan to deploy our framework on a wider selection of embedded
networks. With all of its components being part of the Wiselib, \ACRONYM{} is
able to run over any Wiselib-enabled network. The TupleStore to store RDF data
on embedded devices is in a very mature state. \LINQ{} supports all the major
operations on RDF data for the efficient in-network execution of SPARQL
(sub-)queries. We currently focus on \SELDA{}, extending the SPARQL parser which is required for the and splitting user queries and assembling external an in-network subqueries.

\bibliographystyle{abbrv}
\bibliography{references-compact}

\end{document}